\documentclass[sigconf,nonacm]{acmart}
\usepackage{multirow}
\usepackage{subcaption}
\usepackage{enumitem}
\usepackage{adjustbox}
\usepackage{algorithm2e}

\AtBeginDocument{%
  }

\setcopyright{acmcopyright}
\copyrightyear{2024}
\acmYear{2024}
\acmDOI{XXXXXXX.XXXXXXX}

\acmConference[WSDM '25]{The 18th ACM International Conference on Web Search and Data Mining}{March 10--14,
  2025}{Hannover, Germany}
 
\acmISBN{978-1-4503-XXXX-X/18/06}

\begin{document}

\title{Autoregressive Generation Strategies for Top-K Sequential Recommendations}


\author{Anna Volodkevich}
\email{volodkanna@yandex.ru}
\affiliation{%
  \institution{Sber AI Lab}
  \city{Moscow}
  \country{Russian Federation}
}
\authornote{Authors contributed equally to the paper}

\author{Danil Gusak}
\email{Danil.Gusak@skoltech.ru}
\affiliation{%
  \institution{Skoltech, HSE University}
  \city{Moscow}
  \country{Russian Federation}
}
\authornotemark[1]

\author{Anton Klenitskiy}
\email{antklen@gmail.com}
\affiliation{%
  \institution{Sber AI Lab}
  \city{Moscow}
  \country{Russian Federation}
}
\authornotemark[1]

\author{Alexey Vasilev}
\email{alexxl.vasilev@yandex.ru}
\affiliation{%
  \institution{Sber AI Lab}
  \city{Moscow}
  \country{Russian Federation}
}

\renewcommand{\shortauthors}{Volodkevich, et al.}


\begin{abstract}
  The goal of modern sequential recommender systems is often formulated in terms of next-item prediction. In this paper, we explore the applicability of generative transformer-based models for the Top-K sequential recommendation task, where the goal is to predict items a user is likely to interact with in the “near future”. 

We explore commonly used autoregressive generation strategies, including greedy decoding, beam search, and temperature sampling, to evaluate their performance for the Top-K sequential recommendation task. In addition, we propose novel \textit{Reciprocal Rank Aggregation (RRA)} and \textit{Relevance Aggregation (RA)} generation strategies based on multi-sequence generation with temperature sampling and subsequent aggregation. 

Experiments on diverse datasets give valuable insights regarding commonly used strategies' applicability and show that suggested approaches improve performance on longer time horizons compared to widely-used Top-K prediction approach and single-sequence autoregressive generation strategies. 

\end{abstract}




\maketitle

\section{Introduction}
\label{sec:intro}
Sequential recommender systems with Transformer-based models are a popular area of research. Such systems aim to leverage the order of user-item interactions in historical data to predict future user interactions. However, previous works that have significantly shaped the direction of research within the field, such as SASRec \cite{kang2018self} and BERT4Rec \cite{sun2019bert4rec} were focused on the next-item prediction task. In this formulation, the goal of the model is to predict only one next item in a user sequence. While it is a valid and useful approach for many applications, such as music services or video streaming, prediction over a longer time horizon draws the attention of the scientific community as well \cite{villatel2018recurrent, pancha2022pinnerformer, kolesnikov2021ttrs, devooght2017long, bacciu2023integrating}. 


Further, we denote the number of recommendations by K and the number of ground truth items for each user by N. The goal of Top-K sequential recommendations is to predict a ranked list of K items a user is likely to interact with in the "near future" \cite{tang2018personalized}. Following \cite{villatel2018recurrent}, we aim to correctly predict exactly N hidden future user's actions. This validation strategy allows to easily compare short-term and longer-term performance of considered approaches.



One option to solve this task is to use the Top-K recommendation strategy commonly applied to the next-item prediction task as in \cite{villatel2018recurrent, devooght2017long}. The first limitation of this approach is that the model is not directly trained to predict several next items, so the training objective and final task are not aligned. Second, in this approach, the model scores all candidates simultaneously and independently. So, similar items may have close scores and dominate the recommendation list.

We propose to use a model trained for the next-item prediction task and modify the inference scheme to suit the Top-K sequential recommendation task better. This approach allows to adapt the widely-used Transformer-based models without training objective modification and use the same model for the next-item prediction and the Top-K sequential recommendation. 

We train the GPT-2 model to predict the next item in a sequence, similar to the popular SASRec model \cite{kang2018self}. After that, we autoregressively generate recommendations item-by-item: at each step, we include already generated items in a sequence to generate the following item. So, the model considers already recommended items at each subsequent step and can adjust predictions accordingly. This way, the model can predict more complex and interdependent recommendation lists; for example, it can generate complementary items. The apparent drawback of autoregressive generation is that it is more computationally expensive because we need to score all items $K$ times.

In our paper, we adopt different generation strategies, commonly used for text generation, namely greedy decoding, temperature sampling, and beam search for the Top-K sequential recommendation task. Besides single sequence continuation generation, it is also possible to generate diverse future sequences (several possible realizations of future user behavior) with the same model and combine them into one final recommendation list. We introduce novel approaches with multi-sequence aggregation to make more robust and powerful recommendations. 
This approach leads to some computational overhead but significantly boosts the performance of recommendations. It acts as an ensemble of models, reducing error accumulation on a longer prediction horizon. The difference with standard ensemble methods is that we sample multiple sequences using the same single model, so we don't need to train several models. As we do not aim to predict the exact user sequence ordered by interaction time, we could aggregate generated sequences to enhance the quality of the recommendations.


In short, the main contributions of this paper are:
\begin{itemize}[topsep=2pt]
\item We evaluate several commonly used autoregressive generation strategies and compare them with the Top-K prediction approach to determine their applicability to the Top-K recommendation task. 
\item We propose novel multi-sequence generation approaches, \textit{Reciprocal
Rank Aggregation} and \textit{Relevance Aggregation}, which are based on the generation of several sequences and subsequent aggregation. 
\item Proposed approaches enhance the performance GPT-2 on Top-K sequential recommendations and could be applied to a wide range of generative models trained for the next-item prediction task.
\end{itemize}

\section{Related Work}
\label{sec:literature}

\subsection{Top-K sequential recommendation task}

Recent works on sequential recommendations often consider short-term predictions and namely next-item prediction task, aiming to predict the item a user is going to interact with next \cite{quadrana2018sequence}.
Current state-of-the-art approaches for this task are based on Transformer models, such as SASRec \cite{kang2018self} and BERT4Rec \cite{sun2019bert4rec}.  In the Top-K and long-term sequential recommendations, the goal is to understand what the user will do at a longer time horizon. Works \cite{devooght2017long, villatel2018recurrent} investigate approaches based on recurrent neural networks for long-term sequential recommendations. The Caser model's \cite{tang2018personalized} authors propose the Top-K sequential recommendation approach with convolutional networks. The PinnerFormer paper \cite{pancha2022pinnerformer} uses a Transformer-based model to predict user interactions over a horizon of several days and introduce the new Dense All Action loss designed specifically for this task. In recent work \cite{bacciu2023integrating}, authors argue for the importance of predicting several next items and also create a specific loss function for the SASRec model.

\subsection{Generative approaches to sequential recommendations}

Generative language models such as T5 \cite{raffel2020exploring}, GPT-2 \cite{radford2019language}, and GPT-3 \cite{brown2020language} show significant progress on a wide range of tasks. Adoption of language models' architectures to a recommendation domain is a promising research direction as stated in \cite{de2021transformers4rec} and additionally confirmed by the popularity of the corresponding Transformers4Rec library, which adopts the latest Transformer architectures for sequential recommendation tasks.  

Some works, like P5 \cite{geng2022recommendation} and GPT4Rec \cite{li2023gpt4rec}, adopt pre-trained language models and consider recommendations as text generation tasks. Unlike this work, we consider sequences of item IDs as input and train the model from scratch.

Recent paper \cite{petrov2023generative} also formulates recommendation as an autoregressive item sequence generation with GPT-2. In contrast to our work, the authors evaluate the model on the next-item prediction task while we consider long-term recommendations. In addition, we explore existing commonly used decoding strategies for text generation from NLP \cite{welleck2020consistency, holtzman2019curious, zarriess2021decoding} and introduce a novel multi-sequence aggregation approach.

Autoregressive generation could also be applied to solve the other related sequential recommendation tasks, such as bundle recommendation \cite{bai2019personalized}, aimed at the generation of a collection of associated products that users consume as a whole under the circumstances, e.g., specific intents.
Recent work \cite{sun2023generative} uses autoregressive generation with a Transformer-based model for next-basket recommendation, which refers to the task of predicting a set of items that a user will purchase in the next session, rather than in the current session. Apart from the different tasks, this work considers the Encoder-Decoder model type while we consider the Decoder-only model.

\section{Proposed approach}
\label{sec:approach}
\RestyleAlgo{ruled}
\SetAlgoVlined

\subsection{Recommendation model}\label{sec:model}

In our work, we evaluate the performance of commonly used generation strategies in a Top-K sequential recommendation task and propose novel multi-sequence aggregation strategies. Those strategies may be applied to any autoregressive model, which returns the next-item probability distribution. Proposed approaches do not affect model training and do not impose additional training costs. In our work, we use GPT-2 to conduct experiments. We motivate the model choice and briefly describe GPT-2 architecture and training process in \ref{sec:backbone}.

\subsection{Generation strategies}\label{sec:generation}

\subsubsection{Top-K prediction approach}
Suppose we have a sequence of user interactions $i_{1:t}=\{i_1,i_2,...i_t\}$, and aim to get the Top-K sequential recommendations. One of the commonly used recommendation strategies with sequential models is to apply the model to a known user's interactions sequence, get the model scores for next-item and recommend the Top-K items with the highest score as is done in \cite{villatel2018recurrent, devooght2017long}.

\subsubsection{Autoregressive generation}

In our work, we examine commonly used generation strategies performance in a Top-K sequential recommendation task and propose novel strategies, based on existing ones. Thus, we provide a brief description of the standard generation strategies below. 

Autoregressive generation is based on the assumption that the probability distribution of the  sequence continuation can be decomposed into the product of conditional distributions: 

\begin{equation}\label{eq:autoregressive}
P(i_{t+1:t+K}|i_{1:t}) = \prod_{k=1}^{K} P(i_{t+k}|i_{1:t+k-1})
\end{equation}

Various sequence generation strategies are used and under development in the NLP area \cite{zarriess2021decoding}. Some methods, such as greedy decoding, beam search, and temperature sampling, may be adopted for sequential recommendation tasks.

\subsubsection{Commonly used autoregressive generation strategies}\label{sec:autoregressive_generation}

Each generation strategy described below is based on the next item probability distribution $P(i | i_{1 : t})$, which is inferred as a softmax over an output vector of the last transformer block multiplied by the transposed embedding matrix \cite{radford2018improving}.

For the recommendation task, one needs to convert the generated sequence into a sorted recommendation list. For the commonly used generation strategies we adopt an intuitive approach, which was also used in \cite{petrov2023generative}, and consider the first item in the generated sequence as the first item in a list, the item in the second position as the second item in the recommendation list, and so on.

\textit{Greedy decoding.} Greedy decoding, or greedy search, selects the most probable item as the next item of a sequence:

$$
i_{t + 1} = argmax_i P(i | i_{1 : t})
$$

\textit{Beam search.} Beam search with B beams, where B is known as the beam width, leaves top-B most probable sequence continuations at each generation step, where sequence continuation probability is calculated by the equation \eqref{eq:autoregressive}.
More details of beam search implementation for NLP tasks and its modifications can be found in \cite{vijayakumar2016diverse} and \cite{graves2012sequence}. 

\textit{Temperature sampling.} Another prominent generation strategy is sampling. Following the sampling strategy, the next item is randomly chosen according to its conditional probability distribution:

$$
i_{t + 1} \sim P(i | i_{1 : t})
$$
Softmax temperature tuning may be used to modify probability distribution $P(i | i_{1 : t})$, e.g., to increase probabilities of the most probable items and lower probabilities of long-tail items. Temperature tuning is widely used in text generation \cite{holtzman2019curious} and could be used in recommendation tasks to balance the recommendation's accuracy and diversity. For input sequence $i_{1:t}$, given the output model scores before softmax (logits) $l_i$ for each item $i$ in the item set $I$ and temperature $T$, the softmax is re-estimated as:

$$
P_T(i | i_{1 : t}) = \frac{exp(l_i / T)}{\sum_{j \in I}{exp(l_j / T)}}
$$

Temperature sampling may be combined with top-k sampling \cite{fan2018hierarchical} or Nucleus sampling \cite{holtzman2019curious}. These approaches cut off long-tail items and redistribute the probability mass among selected top items. Following the generation strategy utilized by the original GPT-2 model \cite{radford2019language}, we applied top-k sampling in our experiments. According to this approach, the k most likely next items are filtered, all other items are discarded, and probabilities for those k items are re-normalized. We refer to the sampling strategy as "top-k sampling" and to the baseline prediction strategy as "Top-K prediction strategy" to prevent ambiguity.

\subsubsection{Proposed multi-sequence aggregation strategies.}\label{sec:agg_approach}

We propose to generate multiple diverse sequences with temperature sampling and aggregate generation results into one recommendation list. For this purpose, we introduce and explore two different aggregation strategies described below. 

The proposed approach is closely connected to ensemble techniques, which are widely known to be effective in machine learning \cite{dietterich2000ensemble, caruana2004ensemble}. The classic ensemble approach consists of training several models and combining their predictions to make the model more robust and outperform single models' quality. With the generative model, we can sample multiple sequences using a single model, which is significantly cheaper in terms of computation.



Both proposed strategies share common parts related to the results aggregation. The general pipeline and common parts are described below:
\begin{enumerate}
\item Given a sequence of user interactions $i_{1:t}$ of length $t$ generate $S$ sequence continuations of length $K$ with the trained autoregressive $model$;
\item \textit{Single-sequence generation and aggregation:} For each sequence continuation, calculate scores for all items, which are referred to as $\mathbf{r^s} \in \mathbb{R}^{I}$ in Algorithms \ref{alg:rra}, \ref{alg:ra}. $I$ is a number of items in the catalog and $s$ is a considered sequence continuation number. The proposed approaches differ in this step, therefore, it will be additionally explained in the following subsections.
\item \textit{Multi-sequence aggregation:} 
 Aggregate scores are obtained from each sequence continuation, so the final relevance is equal to the sum of previously assigned scores $\mathbf{r} = \sum_{s=1}^S \mathbf{r^s}$. The $K$ most relevant items from $\mathbf{r}$ constitute the user's recommendations.
\end{enumerate}



\paragraph{Reciprocal Rank Aggregation strategy (RRA)}

The first proposed aggregation strategy is \textit{Reciprocal Rank Aggregation}, adapted from the popular Reciprocal Rank method from Information Retrieval \cite{Wegmann2018SearchFA, Macdonald2006VotingFC, Robertson1995OkapiAT}. 

RRA strategy is illustrated with pseudocode in Algorithm \ref{alg:rra}. Single-sequence generation and aggregation of scores with RRA strategy consists of the following steps:
\begin{enumerate}
\item Generate sequence continuations of length $K$ with temperature sampling. Repetition of already generated items at subsequent generation steps is prohibited.
\item Each item in the generated sequence is assigned relevance equal to an item's reciprocal position (i.e. rank), which gives a relevance sequence of (1, 1/2, 1/3, ...), the relevance of the remaining items is 0.
\end{enumerate}
Thus, we promote items from the beginning of generated sequences to remain in recommendations and simultaneously allow items that frequently appear in continuation to be present in final recommendations. 

\SetAlgoVlined
\begin{algorithm}[htb]
\setlength{\abovecaptionskip}{2pt}
\setlength{\belowcaptionskip}{-7pt}

$\mathbf{r} \gets [0] \times I$\;

\For(\tcp*[f]{computes in parallel}){$s \gets 1$ \KwTo $S$}{

\small
$\mathbf{r^s} \gets [0] \times I$\;

\For{$k \gets 1$ \KwTo $K$}{
$i_{t+k} \gets model.generate(i_{1:t+k-1})$\;
$i_{1:t+k} \gets i_{1:t+k-1}$ $append$ $i_{t+k}$\;
$\mathbf{r^s}_{i_{t+k}} \gets \frac{1}{k}$\;
}
$\mathbf{r} \gets \mathbf{r} + \mathbf{r^s}$;
}
\KwRet{$argsort(-\mathbf{r})[:K]$}\tcp*[r]{get Top-K} 

\caption{Top-K sequential recommendations generation with the RRA strategy} 
\label{alg:rra}
\end{algorithm}


\paragraph{Relevance Aggregation strategy (RA)}

The second aggregation strategy we consider is based on the CombSUM method \cite{Oliveira2020IsRA, Bachanowski2023ACS}. According to it, we aggregate predicted scores (relevances) for each item from each generation step instead of reciprocal ranks of generated items. RA strategy is illustrated with pseudocode in Algorithm \ref{alg:ra}. Single-sequence generation and scores aggregation with RA strategy consists of the following steps:
\begin{enumerate}
\item Generate sequence continuations of length $K$ with temperature sampling. This time we allow the model to predict already generated items and keep predicted relevances $\mathbf{r^{sk}} \in \mathbb{R}^{I}$ for all items on each generation step $k$. Also, for this approach, we don't use top-k sampling.
\item Each item in the catalog is assigned relevance equal to the sum of its relevances at each generation step.
\end{enumerate}
This way, we don't promote items from the beginning of generated sequences and consider all generated steps equivalent.

\SetAlgoVlined
\begin{algorithm}[htb]
\setlength{\abovecaptionskip}{2pt}
\setlength{\belowcaptionskip}{-7pt}
\small
$\mathbf{r} \gets [0] \times I$\;

\For(\tcp*[f]{computes in parallel}){$s \gets 1$ \KwTo $S$}{
$\mathbf{r^s} \gets [0] \times I$\;

\For{$k \gets 1$ \KwTo $K$}{
$i_{t+k}$, $r^{sk} \gets model.generate(i_{1:t+k-1})$\;
$i_{1:t+k} \gets i_{1:t+k-1}$ $append$ $i_{t+k}$\;
$\mathbf{r^s} \gets \mathbf{r^s} + \mathbf{r^{sk}}$\;
}
$\mathbf{r} \gets \mathbf{r} + \mathbf{r^s}$;
}
\KwRet{$argsort(-\mathbf{r})[:K]$}\tcp*[r]{get Top-K} 

\caption{Top-K sequential recommendations generation with the RA strategy} 
\label{alg:ra}
\end{algorithm}





\section{Experiments}
\label{sec:experiments}
In this section, we present our experimental setup and
empirical results. Our experiments are designed to answer the following research questions:
\begin{description}
\item[\textbf{RQ1:}] How well do the autoregressive generation strategies, such as greedy and beam search, perform in the Top-K sequential recommendation task compared to the Top-K prediction approach?
\item[\textbf{RQ2:}] Do the proposed multi-sequence aggregation strategies outperform the autoregressive generation strategies? How do the hyperparameters affect their performance?  
\item[\textbf{RQ3:}] How significant the computational overhead of multi-sequence aggregation is compared to standard generation approaches? 
\item[\textbf{RQ4:}] How well do all the considered strategies perform on a longer prediction horizon? Do the proposed multi-sequence aggregation strategies perform better in predicting longer-term user preferences?
\end{description}

\subsection{Experimental Settings}
\label{sec:experimental_settings}

\subsubsection{Datasets}\label{sec:datasets}


We conduct experiments on six datasets collected from real-world platforms. The datasets vary significantly in domains, sparsity, and types of feedback:
\begin{itemize}
\item  \textbf{MovieLens-20M (ML-20M) \cite{harper2015movielens}:} This is a movie recommendation dataset, which is widely used for benchmarking sequential recommenders \cite{wu2021ml, cho2020ml, zhao2020ml, li2021ml, fischer2020ml}. In this work, ML-20M is the dataset with the largest number of user-item interactions after preprocessing.
\item  \textbf{Yelp \cite{asghar2016yelp}:} This is a business reviews dataset and is another popular benchmark for sequential recommendations \cite{padungkiatwattana2022yelp, amjadi2021yelp, ruihong2021yelp}, and is the sparsest dataset in the list.
\item  \textbf{Steam \cite{pathak2017steam}:} This is a dataset collected from Steam, a large online video game distribution platform.
\item  \textbf{Gowalla \cite{cho2011gowalla}:} The dataset from Gowalla – location-based social networking platform, where users share their locations by checking in.
\item  \textbf{Twitch-100k \cite{rappaz2021twitch}:} This is a dataset of 100 thousand users consuming streaming content on Twitch.
\item  \textbf{BeerAdvocate \cite{mcauley2012beer}:} This dataset consists of beer reviews from BeerAdvocate and has the highest average interaction length of all datasets.
\end{itemize}

As in previous publications \cite{tang2018personalized, kang2018self, sun2019bert4rec}, the presence of a review or rating is considered implicit feedback.
For testing a model's ability to make long-term recommendations, it's essential to use datasets with a long enough history of interactions for each user. We filter out users with fewer than 20 interaction records for these purposes. Following common practice \cite{zhang2019feature, rendle2010factorizing}, we also discard unpopular items with less than five interactions. The final statistics of datasets are summarized in Table \ref{tab:datasetStats}.


\begin{table}[htb]
\setlength{\abovecaptionskip}{7pt}
\caption{\textbf{{Statistics of the datasets after preprocessing.}}} \label{tab:datasetStats}
\resizebox{\columnwidth}{!}{%
\begin{tabular}{lrrrrr}
\hline
\textbf{Dataset} & \textbf{\#Users} & \textbf{\#Items} & \textbf{\#Interactions} & \textbf{Avg. length} & \textbf{Density} \\ \hline
ML-20M       & 138,476 & 18,345  & 19,983,706 & 144.3 & 0.79\% \\
Yelp         & 42,429  & 137,039 & 2,294,804  & 54.1  & 0.04\% \\
Steam        & 33,375  & 11,945  & 1,448,441  & 43.4  & 0.36\% \\
Gowalla      & 27,516  & 173,511 & 2,627,663  & 95.5  & 0.06\% \\
Twitch-100k  & 20,899  & 28,339  & 1,577,168  & 75.5  & 0.27\% \\
BeerAdvocate & 7,606   & 22,307  & 1,409,494  & 185.3 & 0.83\% \\ \hline
\end{tabular}%
}
\vspace{-10pt}
\end{table}

\subsubsection{Evaluation}\label{sec:evals}

We applied the following recommendation list generation constraints: we do not generate items already presented in the user history and forbid items repetition in the generated sequence (except for the singe-sequence generation step in RA approach, see section \ref{sec:agg_approach}).

We follow \cite{villatel2018recurrent} and hold the last $N$ items of each user interaction sequence for validation and testing with $N=10$. All previous interactions are used for training. Data with the last $N$ items is further split into validation and test set randomly by users. Hyperparameter tuning is performed on the validation set, and the final performance is evaluated on the test set.

For performance evaluation, we use three commonly used ranking metrics: Normalized Discounted Cumulative Gain (NDCG@10), Recall@10, and Mean Average Precision (MAP@10) from the popular library Recommenders\footnote{\url{https://github.com/recommenders-team/recommenders}}. In addition, we look at the performance for each position in a test sequence separately. In this case, ground truth consists of only one item on the corresponding position in the original ground truth sequence. Such an approach was used in \cite{villatel2018recurrent}. This is done to more thoroughly evaluate how generation strategy influences predictions on longer-term horizons.

\subsubsection{Generative model (GPT-2)}\label{sec:backbone}
We use GPT-2 \cite{radford2019language} model in our experiments as it is a popular generative model architecturally close to SASRec \cite{petrov2023generative}. To be able to use a wide range of sequence generation methods (decoding strategies) out of the box\footnote{\url{https://huggingface.co/blog/how-to-generate}} we utilized its implementation in HuggingFace Transformers library \cite{wolf2019huggingface}. 

We train GPT-2 from scratch using standard language modeling objective (with item IDs instead of token IDs as inputs). The model aims to predict the input sequence shifted by one item to the left. Item probabilities are modeled with softmax operation over model outputs. The model is trained with the cross-entropy loss, which corresponds to the maximum likelihood principle, and is used in language modeling and many sequential recommendations models \cite{sun2019bert4rec, klenitskiy2023turning, petrov2023generative}.

\subsubsection{Baseline methods}\label{sec:baselines}

The proposed generation strategies may be applied to any autoregressive model, which can return the next-item probability distribution. Thus, our main goal is to compare suggested generation approaches with standard prediction approach without generation. According to this, our primary baseline for comparison is the same \textbf{GPT-2} model with a standard Top-K prediction strategy. 

Additional baselines are \textbf{BPR-MF} (classic matrix factorization-based approach with a
pairwise BPR loss) and two the most widely used state-of-the-art baselines in sequential recommendations - \textbf{SASRec} and \textbf{BERT4Rec}. In recent papers \cite{klenitskiy2023turning, wilm2023scaling, petrov2023gsasrec}, it was shown that the SASRec model could achieve much better performance if trained with the cross-entropy loss or sampled cross-entropy loss instead of original binary cross-entropy loss. So we train SASRec with the cross-entropy loss and refer to it as \textbf{SASRec+} following \cite{klenitskiy2023turning}.

\subsubsection{Implementation Details}\label{sec:details}

We choose the hidden representation size to be equal to 256 for MovieLens-20M and 64 for all other datasets based on the results of hyperparameter tuning.
For all models, we use a dropout rate of 0.1, two self-attention blocks, and one attention head (except for two attention heads for BERT4Rec). The masking probability for BERT4Rec was set to 0.2. These settings are consistent with parameters used in previous papers \cite{kang2018self, sun2019bert4rec, petrov2022systematic}. We set a maximum sequence length of 128. All models were trained with a batch size of 64 and Adam optimizer with a learning rate 1e-3. We use the early stopping criterion to determine the number of training epochs.
For temperature sampling and Reciprocal Rank Aggregation strategy, there is an additional parameter top-k to cut off long-tail items. We set its value to 10 as we find this to work the best experimentally (see section \ref{sec:choosing_topk}).
For BPR-MF, we use fast GPU implementation from the Implicit library \cite{frederickson2018fast} and tuned the number of latent components, regularization, and the learning rate with Optuna \cite{akiba2019optuna}.
The code to reproduce experiments is provided in the corresponding GitHub repository\footnote{\url{https://github.com/dalibra/autoregressive-generation-strategies-longterm-seqrec}}.
HuggingFace Transformers sequence generation functionality was used in this code. All experiments were conducted using an NVIDIA Tesla V100 GPU.

\subsection{Single Sequence Generation}\label{sec:beam_temp}

We tuned the main parameters for beam search and temperature sampling (number of beams and temperature respectively) and found that none of these strategies outperform simple greedy decoding. This is expected for temperature sampling, as sampling means taking more random and less optimal predictions. However, it is less expected for beam search and contradicts experience from text generation. We analyze these results in the following sections.

\subsubsection{Temperature sampling.}\label{sec:temp_sampling}

Figure \ref{fig:temp} demonstrates performance for different temperature values. A low-temperature setting is effectively equivalent to greedy decoding. As the temperature increases, model performance degrades. This degradation occurs because sampling at higher temperatures introduces a greater degree of randomness into the prediction process. Instead of prioritizing the most probable item, the model selects items with lower confidence levels at each step, leading to a decline in performance metrics. Despite this drawback, sampling with high enough temperatures can enhance the diversity recommendations and plays an important role in multi-sequence aggregation strategies.

\subsubsection{Beam search.}\label{sec:beam_search}

\begin{figure*}[htbp]
    \centering
    \begin{minipage}{0.49\textwidth}
        \setlength{\abovecaptionskip}{2pt}
        \setlength{\belowcaptionskip}{-7pt}
        \centering
        \includegraphics[width=0.9\textwidth]{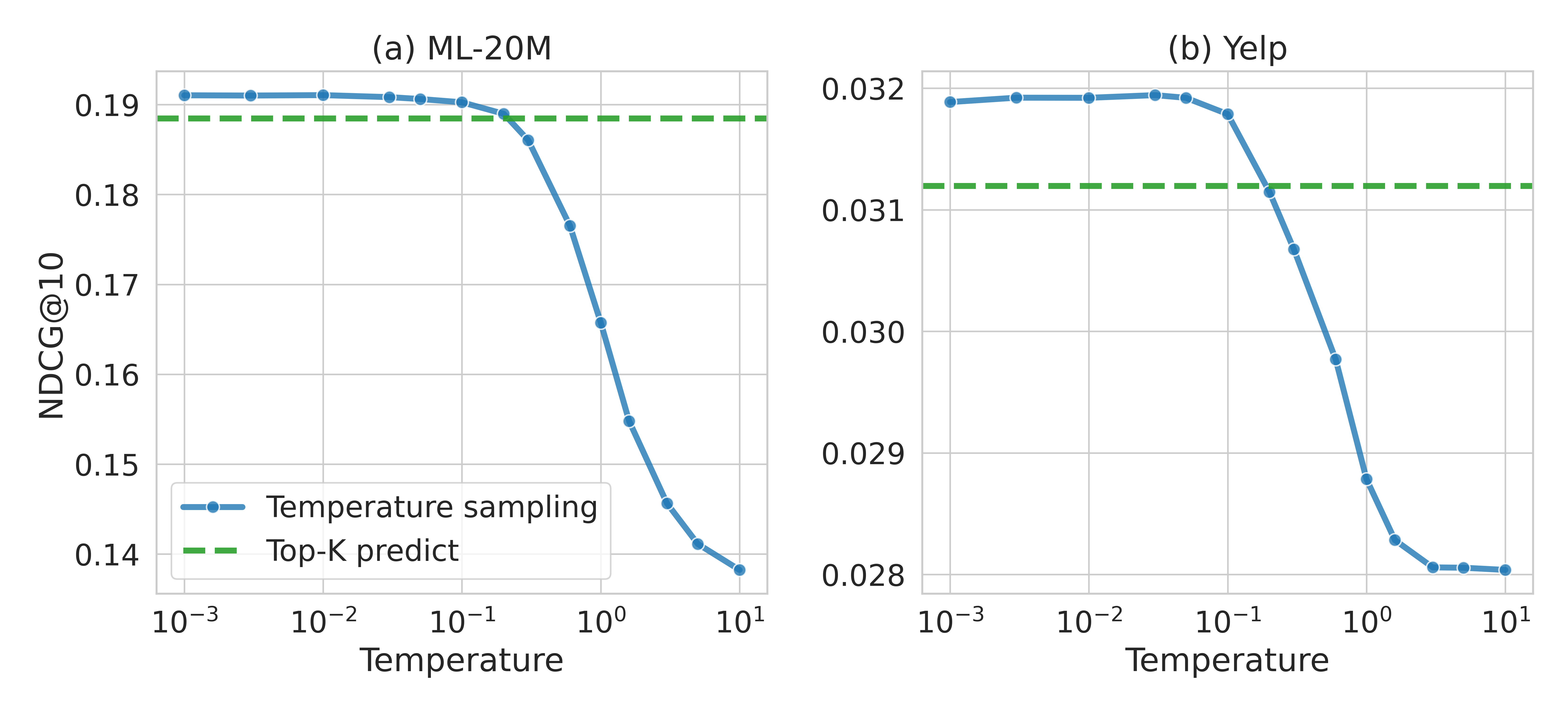}
        \subcaption{Temperature sampling. The left point is close to greedy decoding.}
        \label{fig:temp}
    \end{minipage}
    \hfill
    \begin{minipage}{0.49\textwidth}
        \setlength{\abovecaptionskip}{2pt}
        \setlength{\belowcaptionskip}{-7pt}
        \centering
        \includegraphics[width=0.9\textwidth]{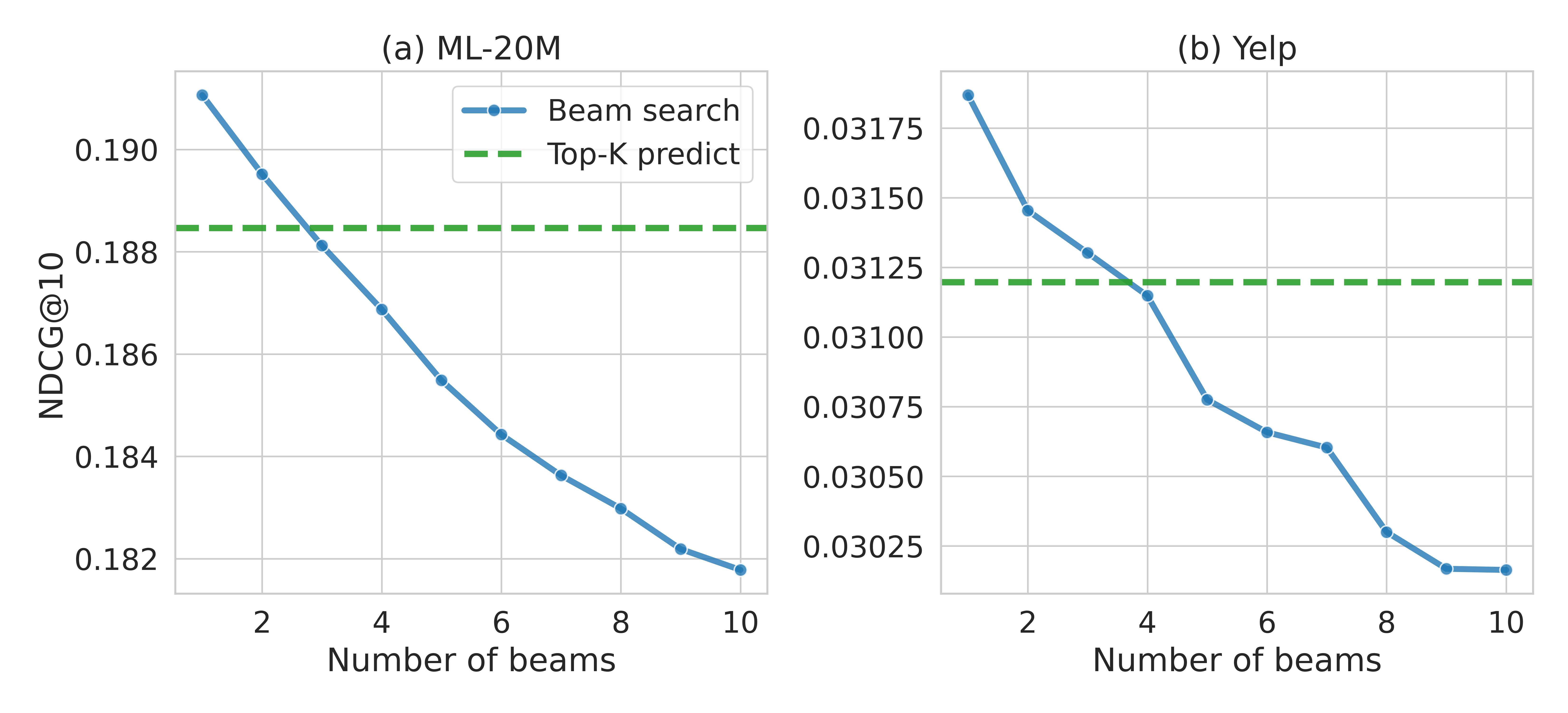}
        \subcaption{Beam search. The leftmost point is equivalent to greedy decoding.}
        \label{fig:num_beams}
    \end{minipage}
    \caption{Comparison of different generation strategies and their impact on NDCG@10 on Movielens-20M and Yelp datasets. The horizontal dashed lines correspond to the Top-K prediction strategy without generation.}
    \Description[Comparison of different generation strategies]{Comparison of different generation strategiesy}
\end{figure*}

The observation that beam search, despite its efficiency in NLP tasks, does not outperform greedy decoding in the field of recommender systems is a less expected result. Figure \ref{fig:num_beams} illustrates that the optimal number of beams is one, which corresponds to greedy generation.

To explain this observation, we note that in text generation the goal is to produce plausible and coherent text. Each step in the sequence is equally important, with the end of the sentence requiring as much accuracy and relevance as the beginning. In the context of recommender systems, the situation is different. Predicting the initial steps is inherently simpler while predicting subsequent steps is significantly more challenging due to an error accumulation. In the process of choosing the optimal beam, beam search considers the probabilities of later items and makes less probable choices for the first steps. While this approach might slightly improve the quality of later steps, early step errors lead to overall performance degradation. With the greedy approach, on the contrary, the most probable items are chosen for the first steps, which leads to better results.

To confirm this explanation, we measure HitRate for each generation step separately in both greedy and beam search scenarios (Table \ref{tab:hr_reversed}). In the first step, the difference is significant, with beam search severely degrading the quality due to choosing less probable items. This gap diminishes in later steps, and towards the sequence's end, beam search performs better.

\begin{table}[ht]
\setlength{\abovecaptionskip}{2pt}
\setlength{\belowcaptionskip}{-1pt}
\caption{\textbf{{Dependence of the HitRate on generation step for Steam dataset.}}} \label{tab:hr_reversed}
\small
\resizebox{\columnwidth}{!}{%
\begin{tabular}{lcccccccccc}
Generation step \#               & 1      & 2      & 3      & 4      & 5      & 6      & 7      & 8      & 9      & 10     \\ \hline
Greedy\textsubscript{GPT-2}      & 0.101 & 0.083 & 0.077 & 0.074 & 0.070 & 0.066 & 0.061 & 0.056 & 0.055 & 0.050 \\
Beam search\textsubscript{GPT-2} & 0.076 & 0.075 & 0.074 & 0.070 & 0.063 & 0.059 & 0.060 & 0.053 & 0.056 & 0.057
\end{tabular}%
}

\end{table}



\subsection{Multi-Sequence Aggregation}\label{sec:aggregation}

\subsubsection{Choosing top-k for reciprocal rank aggregation.}\label{sec:choosing_topk} 
\begin{figure}[htb]
    \centering
        \includegraphics[width=0.40\textwidth]{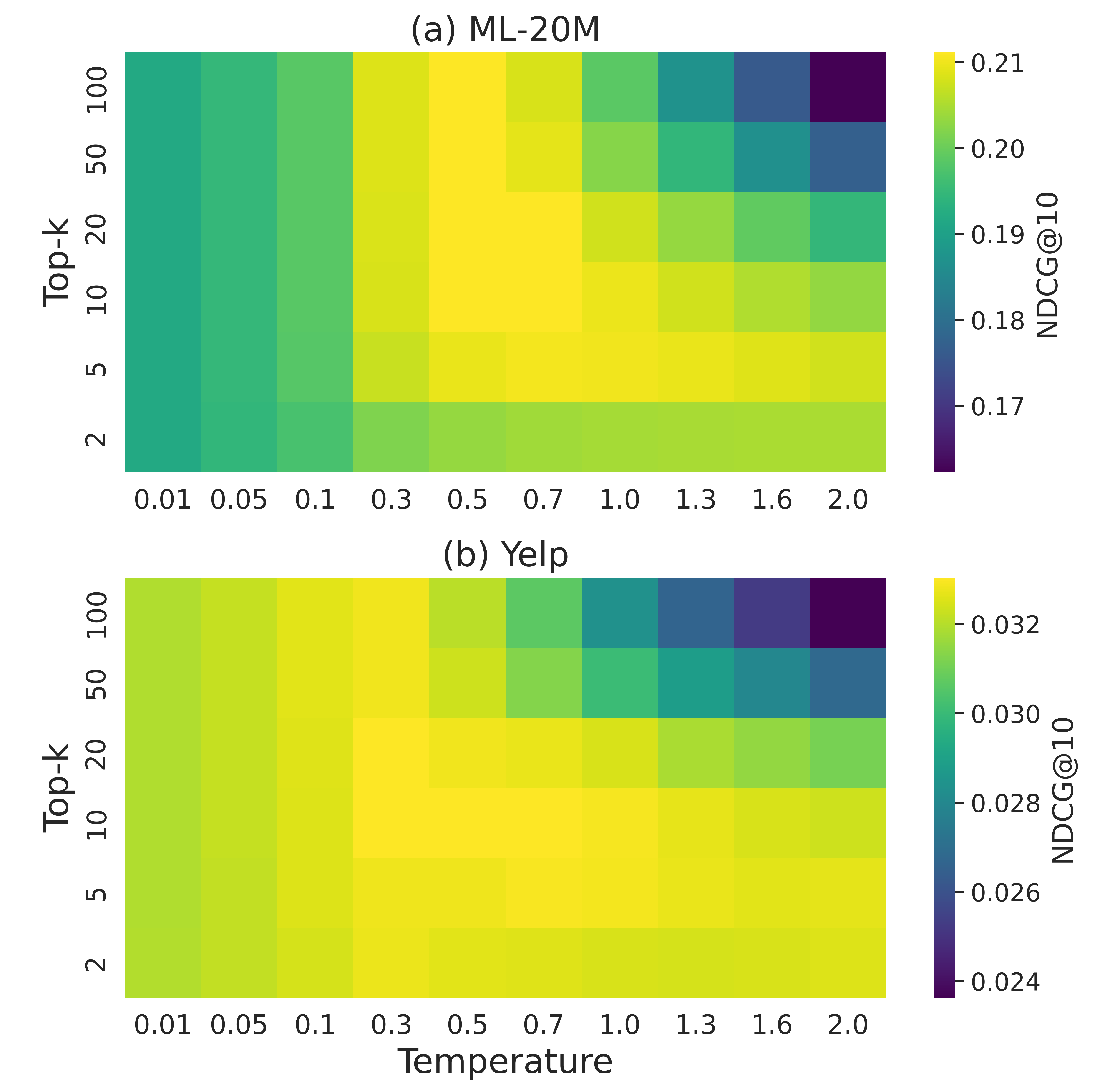}
        \caption{NDCG@10 for different top-k and temperature values. Reciprocal Rank Aggregation with 30 sequences.}
        \Description[NDCG for different top-k and temperature values]{NDCG for different top-k and temperature values}
        \label{fig:top_k}
\end{figure}

When we use temperature sampling and reciprocal rank aggregation, there is an additional hyperparameter to tune - top-k (see section \ref{sec:autoregressive_generation}). The best top-k value can depend on temperature. So, we perform a grid search for top-k and temperature for reciprocal rank aggregation with 30 sequences on two datasets, MovieLens-20M and Yelp. The results are shown in Figure \ref{fig:top_k}. For low-temperature performance for different top-k values is the same. With lower temperatures, only the most relevant items are generated, which makes the generated sequences more similar, and thus, aggregation of those sequences does not allow to to achieve competitive quality. But for higher temperature values, dependence on top-k becomes significant. With high top-k values (or without top-k sampling), the model generates more irrelevant items which leads to a decrease in quality. We choose top-k=10 for all experiments with the Reciprocal Rank Aggregation strategy based on the grid search results.



\subsubsection{Dependency on temperature value.}\label{sec:temp_dependency} 
\begin{figure*}
\setlength{\abovecaptionskip}{2pt}
\includegraphics[width=0.8\textwidth]{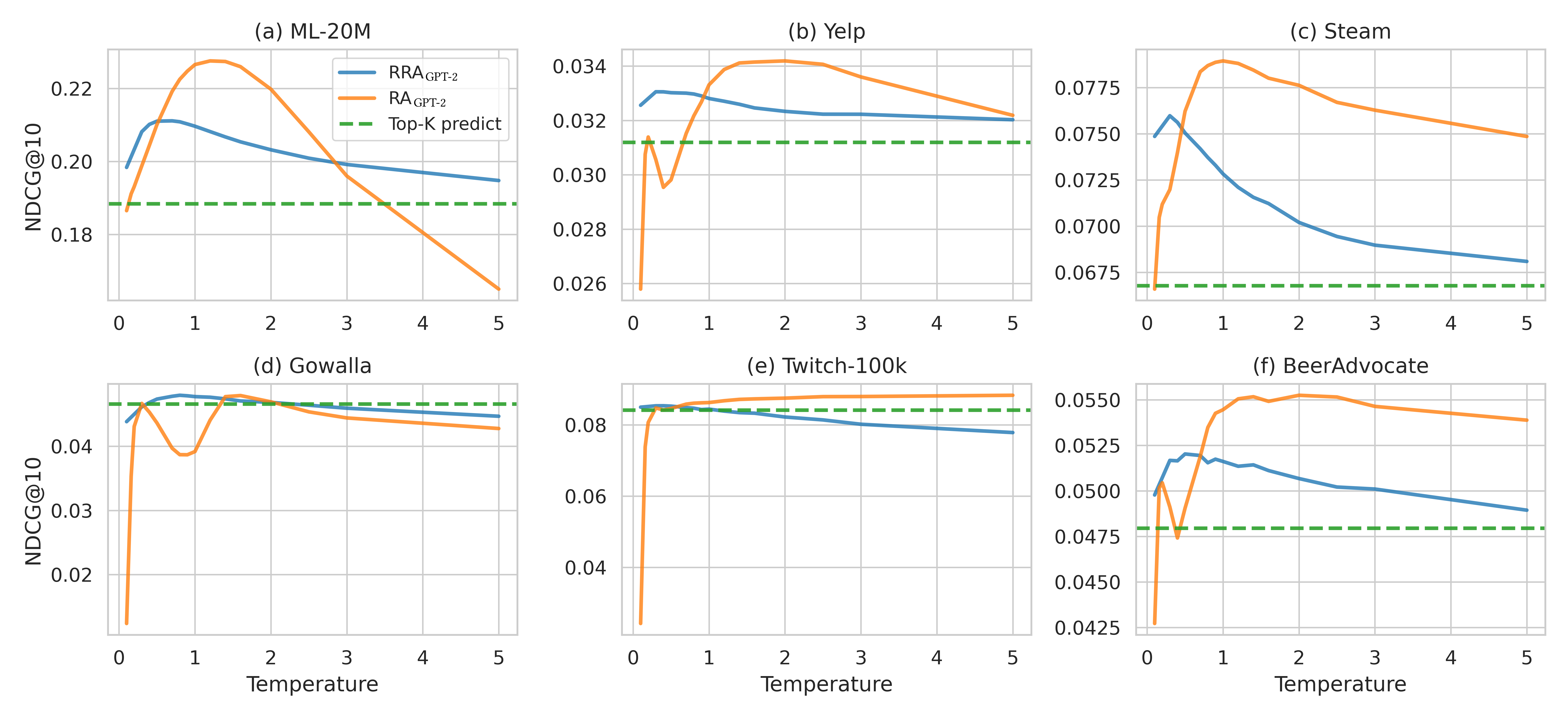}
\caption{NDCG@10 for different values of temperature. Multi-sequence aggregation with 30 sequences on all datasets. The horizontal dashed line corresponds to the Top-K prediction strategy without generation.}
\Description[NDCG for different values of temperature]{NDCG for different values of temperature}
\captionsetup{justification=centering}
\label{fig:temp_agg}
\end{figure*}

When we aggregate multiple sequences, sampling with temperature turns from weakness to strength. To take advantage of combining several predictions, we need to generate diverse enough sequences, thus, relatively high temperature values are desirable. Figure \ref{fig:temp_agg} demonstrates NDCG@10 for different temperature values for proposed aggregation strategies on all datasets. We chose 30 sequences for aggregation as a reasonable trade-off between quality and speed. As shown in section  \ref{sec:num_seq_dependency}, taking more sequences won't result in significant improvements.

For MovieLens-20M, Yelp, BeerAdvocate, and Steam, both aggregation strategies consistently outperform standard Top-K prediction for a wide range of temperature values. There is not enough diversity for too low temperature, while for too high temperature, there is too much randomness, so proper tuning of this parameter is essential. The Relevance Aggregation strategy shows better results than the Reciprocal Rank Aggregation on all datasets except for Gowalla. 

\subsubsection{Dependency on the number of sequences.}\label{sec:num_seq_dependency} 
\begin{figure*}[htbp!]
    \centering
    \setlength{\abovecaptionskip}{5pt} 
    \setlength{\belowcaptionskip}{0pt} 

    \begin{minipage}{0.49\textwidth}
        \centering
        \includegraphics[width=0.9\textwidth,height=5cm,keepaspectratio]{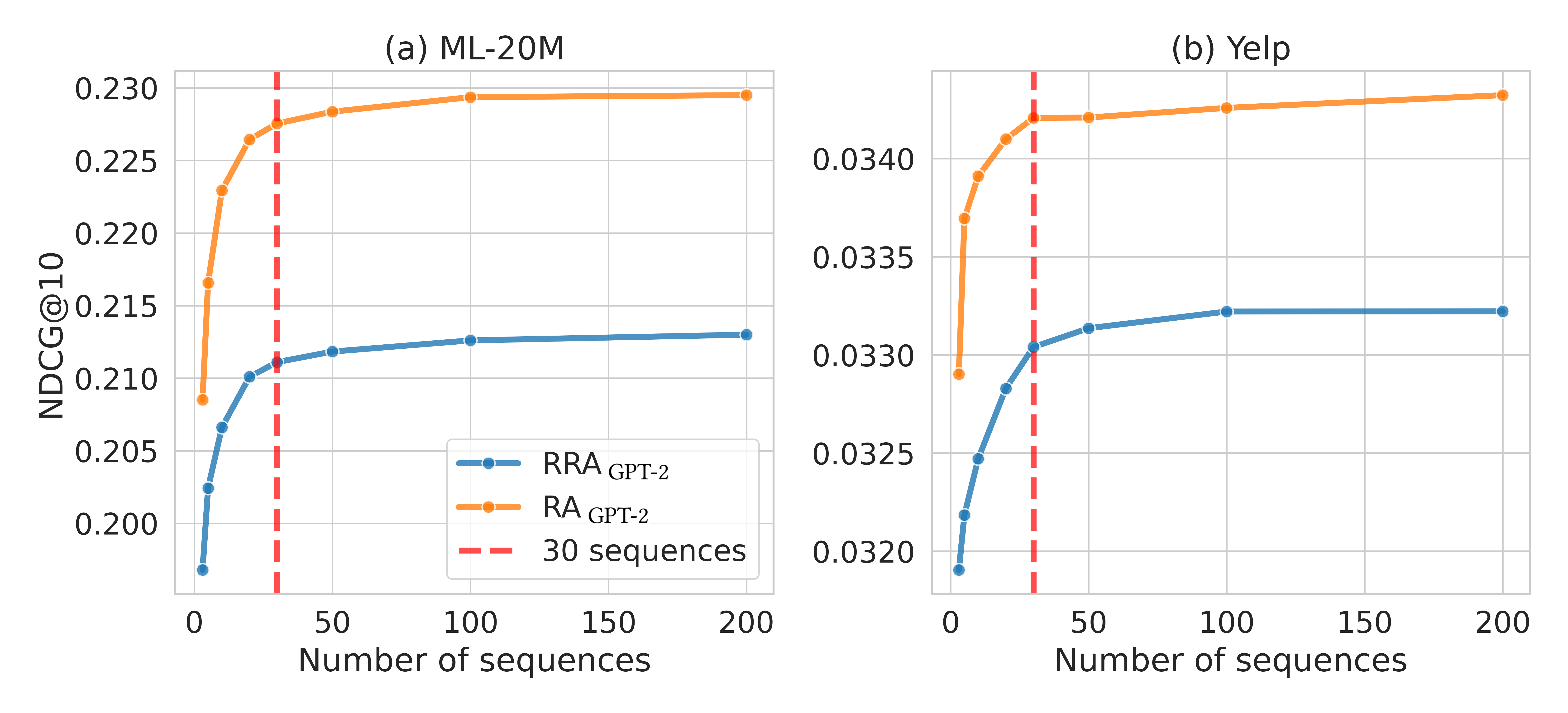}
        \subcaption{NDCG@10 for a different number of sequences.}
        \vspace{0.42cm} 
        \label{fig:num_seq}
    \end{minipage}%
    \hfill
    \begin{minipage}{0.49\textwidth}
        \centering
        \includegraphics[width=0.9\textwidth,height=5cm,keepaspectratio]{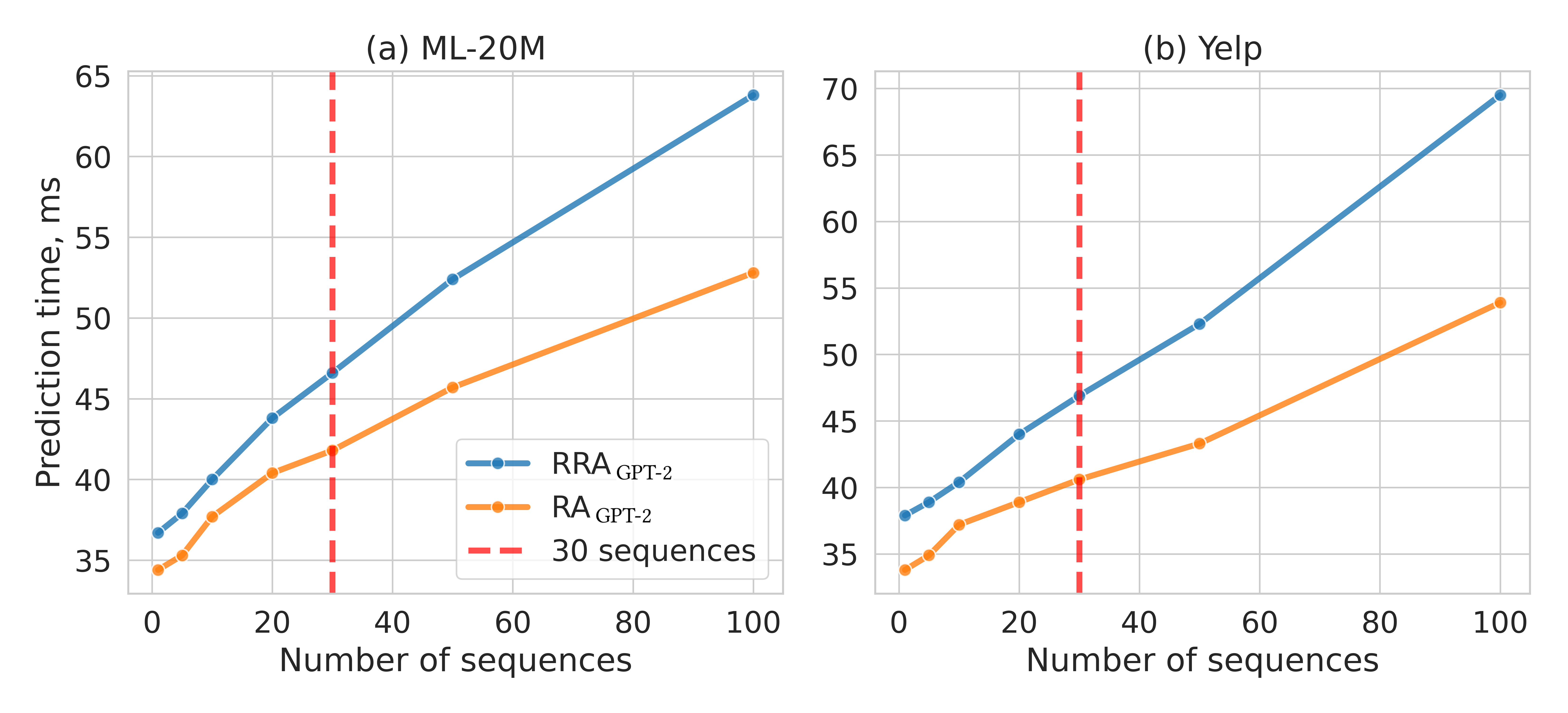}
        \subcaption{Prediction time for a single user for different numbers of sequences in milliseconds.}
        \label{fig:time}
    \end{minipage}

    \caption{Comparative analysis of different multi-sequence aggregation strategies with the best temperature on Movielens-20M and Yelp datasets. The vertical dashed line marks the number of sequences equal to 30, which was used in other experiments.}
    \Description[Comparative analysis of different multi-sequence aggregation strategies]{Comparative analysis of different multi-sequence aggregation strategies}
\end{figure*}

Figure \ref{fig:num_seq} demonstrates NDCG@10 for different number of sequences to aggregate on MovieLens-20M and Yelp datasets. The results for both aggregation strategies are very similar. Metrics are constantly growing with the addition of sequences, but most of the improvement is achieved in the beginning. This result is expected because aggregating multiple sequences acts are similar to the models' ensemble. Around 30 sequences, performance saturates and starts to grow slower, so we fixed this number of sequences for the other experiments. Certainly, there is a trade-off between performance improvement and the additional computational cost of adding more sequences. Aggregating too many sequences will not be practically useful.


\subsubsection{Inference speed considerations.}\label{sec:speed} 

Generating multiple sequences with subsequent aggregation brings additional computation costs compared with the baseline Top-K prediction approach. We report an inference time for the entire set of users on Movielens-20M and Yelp datasets in Table \ref{tab:predictionTimes}.  The computational overhead of aggregation strategies for the entire test set with fixed GPU resources is significant.

However, in real-world systems, inference is highly parallelized by users. Besides, multiple sequence generation could be performed in parallel for a single user, so the latency overhead is manageable and could be further reduced. We analyzed the inference time per user based on the number of generated sequences (see Figure \ref{fig:time}) and confirmed that the dependence is close to linear with a relatively slight slope. The inference time for 30 sequences is only 1.5 times longer than the inference time for one sequence. In addition, the inference time per user for 30 sequences is less than 50 ms, which demonstrates the efficiency of the proposed approach in terms of latency. This paves the way for further refinement and potential deployment of multi-sequence aggregation strategies in industrial applications. 


\begin{table}[htb]
\setlength{\abovecaptionskip}{2pt}
\setlength{\belowcaptionskip}{-1pt}
\caption{\textbf{{Prediction time in seconds for entire set of users on Movielens-20M and Yelp datasets. For the Reciprocal Rank Aggregation and Relevance Aggregation strategies, the number of sequences equals 30. The test batch sizes for the Movielens-20M and Yelp are set at 72 and 12, respectively.}}} \label{tab:predictionTimes}
\small
\resizebox{0.8\columnwidth}{!}{%
\begin{tabular}{l|c|ccc}
\hline
Dataset & GPT-2 & Greedy\textsubscript{GPT-2} & RRA\textsubscript{GPT-2} & RA\textsubscript{GPT-2}\\ \hline
ML-20M     & 32 & 89        & 1281         & 874         \\
Yelp      & 31  & 122        & 709          & 570          \\ \hline
\end{tabular}%
}
\vspace{-10pt}
\end{table}

\begin{table*}[htb]
\setlength{\abovecaptionskip}{7pt}
\caption{\textbf{Performance comparison on all datasets. Bold scores are the best in the baselines/proposed strategies subgroups, while underlined scores are the second best. The last column represents the relative improvement of the best-proposed strategy over the best baseline. The symbol * denotes the statistical significance via paired t-test at $p < 0.05$ when comparing the proposed strategy performance with the best baseline. GPT-2 stands for Top-K prediction strategy, Greedy\textsubscript{GPT-2} - greedy decoding, RRA\textsubscript{GPT-2} - Reciprocal Rank Aggregation, RA\textsubscript{GPT-2}- Relevance Aggregation.}} 
\label{tab:perf_comp}
\resizebox{0.8\textwidth}{!}{
\begin{tabular}{|l|r|llll|lll|r|}
\hline
Dataset &
  Metric &
  BPR-MF &
  BERT4Rec &
  SASRec+ &
  GPT-2 &
  Greedy\textsubscript{GPT-2} &
  RRA\textsubscript{GPT-2} &
  RA\textsubscript{GPT-2}&
  Improv. \\ \hline
\multirow{3}{*}{ML-20M} &
  NDCG@10 &
  0.0732 &
  0.1713 &
  \underline{0.1852} &
  \textbf{0.1876} &
  0.1897* &
  \underline{0.2103*} &
  \textbf{0.2280*} &
  21.54\% \\
 &
  Recall@10 &
  0.0682 &
  0.1553 &
  \underline{0.1677} &
  \textbf{0.1693} &
  0.1691 &
  \underline{0.1933*} &
  \textbf{0.2073*} &
  22.45\% \\
 &
  MAP@10 &
  0.0313 &
  0.0872 &
  \underline{0.0961} &
  \textbf{0.0979} &
  0.1055* &
  \underline{0.1165*} &
  \textbf{0.1277*} &
  30.44\% \\ \hline
\multirow{3}{*}{Yelp} &
  NDCG@10 &
  0.0179 &
  0.0268 &
  \underline{0.0301} &
  \textbf{0.0305} &
  0.0310* &
  \underline {0.0323*} &
  \textbf{0.0339*} &
  11.15\% \\
 &
  Recall@10 &
  0.0161 &
  0.0253 &
  \underline{0.0283} &
  \textbf{0.0286} &
  0.0293* &
  \underline{0.0304*} &
  \textbf{0.0315*} &
  10.14\% \\
 &
  MAP@10 &
  0.0058 &
  0.0091 &
  \underline{0.0105} &
  \textbf{0.0106} &
  0.0109* &
  \underline{0.0113*} &
  \textbf{0.0119*} &
  12.26\% \\ \hline
\multirow{3}{*}{Steam} &
  NDCG@10 &
  0.0434 &
  \textbf{0.0689} &
  \underline{0.0682} &
  0.0657 &
  0.0706* &
  \underline{0.0743*} &
  \textbf{0.0777*} &
  12.77\% \\
 &
  Recall@10 &
  0.0395 &
  \textbf{0.0621} &
  \underline{0.0608} &
  0.0586 &
  0.0644* &
  \underline{0.0681*} &
  \textbf{0.0698*} &
  12.40\% \\
 &
  MAP@10 &
  0.0156 &
  \textbf{0.0258} &
  \underline{0.0257} &
  0.0248 &
  0.0271* &
  \underline{0.0287*} &
  \textbf{0.0303*} &
  17.44\% \\ \hline
\multirow{3}{*}{Gowalla} &
  NDCG@10 &
  0.0119 &
  0.0296 &
  \underline{0.0470} &
  \textbf{0.0471} &
  0.0430 &
  \underline{0.0477} &
  \textbf{0.0480*} &
  1.91\% \\
 &
  Recall@10 &
  0.0085 &
  0.0260 &
  \underline{0.0402} &
  \textbf{0.0404} &
  0.0355 &
  \underline{0.0411*} &
  \textbf{0.0414*} &
  2.48\% \\
 &
  MAP@10 &
  0.0033 &
  0.0122 &
  \underline{0.0207} &
  \textbf{0.0209} &
  0.0195 &
  \textbf{0.0210} &
  \underline{0.0207} &
  0.48\% \\ \hline
\multirow{3}{*}{Twitch-100k} &
  NDCG@10 &
  0.0822 &
  \underline{0.0849} &
  \textbf{0.0853} &
  0.0847 &
  0.0852 &
  \underline{0.0863} &
  \textbf{0.0882*} &
  3.40\% \\
 &
  Recall@10 &
  0.0514 &
  \underline{0.0726} &
  \textbf{0.0728} &
  0.0721 &
  0.0732 &
  \underline{0.0740*} &
  \textbf{0.0745*} &
  2.34\% \\
 &
  MAP@10 &
  0.0275 &
  \textbf{0.0401} &
  \underline{0.0397} &
  0.0392 &
  0.0394 &
  \underline{0.0399} &
  \textbf{0.0416*} &
  3.74\% \\ \hline
\multirow{3}{*}{BeerAdvocate} & NDCG@10 & 0.0358 & 0.0407 & \textbf{0.0514} & \underline{0.0498} & 0.0476 & \underline{0.0537*} & \textbf{0.0569*} & 10.70\% \\
 &
  Recall@10 &
  0.0328 &
  0.0374 &
  \textbf{0.0465} &
  \underline{0.0451} &
  0.0421 &
  \underline{0.0484*} &
  \textbf{0.0524*} &
  12.69\% \\
 &
  MAP@10 &
  0.0128 &
  0.0147 &
  \textbf{0.0200} &
  \underline{0.0190} &
  0.0184 &
  \underline {0.0205*} &
  \textbf{0.0212*} &
  6.00\% \\ \hline
\end{tabular}%
}
\end{table*}

\subsection{Overall Performance}\label{sec:performance}

Table \ref{tab:perf_comp} summarizes the main results of experiments on all datasets. We compare the baseline autoregressive generation strategy (greedy decoding) with the two proposed multi-sequence aggregation strategies (Reciprocal Rank Aggregation and Relevance Aggregation) and the baselines from section \ref{sec:baselines}. We don't include beam search and temperature sampling in the final comparison because optimal parameters for these strategies correspond to greedy decoding (number of beams equals one and the lowest considered temperature); see section \ref{sec:beam_temp}.

For multi-sequence aggregation, 30 sequences and the best value of temperature for each dataset are used (for reciprocal rank aggregation and relevance aggregation respectively: 0.5 and 1.2 for Movielens-20M, 0.5 and 1.8 for Yelp, 0.3 and 1.0 for Steam, 0.8 and 1.6 for Gowalla, 0.4 and 5.0 for Twitch-100k, 0.5 and 2.0 for BeerAdvocate).

Generation with greedy decoding is comparable with the standard Top-K prediction strategy. It is slightly better on some datasets (MovieLens-20M, Yelp, Steam) and a little worse on others (Gowalla, BeerAdvocate). However, as shown in section \ref{sec:time_step}, it could be better at predicting later time steps. With multi-sequence aggregation, it is possible to significantly outperform baselines on 4 of 6 datasets (MovieLens-20M, Yelp, BeerAdvocate, and Steam). For Gowalla and Twitch-100k improvements are very small. The relevance aggregation strategy is consistently better than reciprocal rank aggregation. To summarize, we noted statistically significant improvement in quality metrics for the majority of considered datasets. Thus, we confirmed that proposed generation strategies could be applied for datasets significantly varying in domains and sparsity. In the next section, the influence on longer time horizons is investigated further.

\subsection{Performance by Ground Truth Item Position}\label{sec:time_step}

\begin{figure}[htbp]
    \centering
    \setlength{\abovecaptionskip}{2pt}
    \setlength{\belowcaptionskip}{-10pt}
    \includegraphics[width=\linewidth]{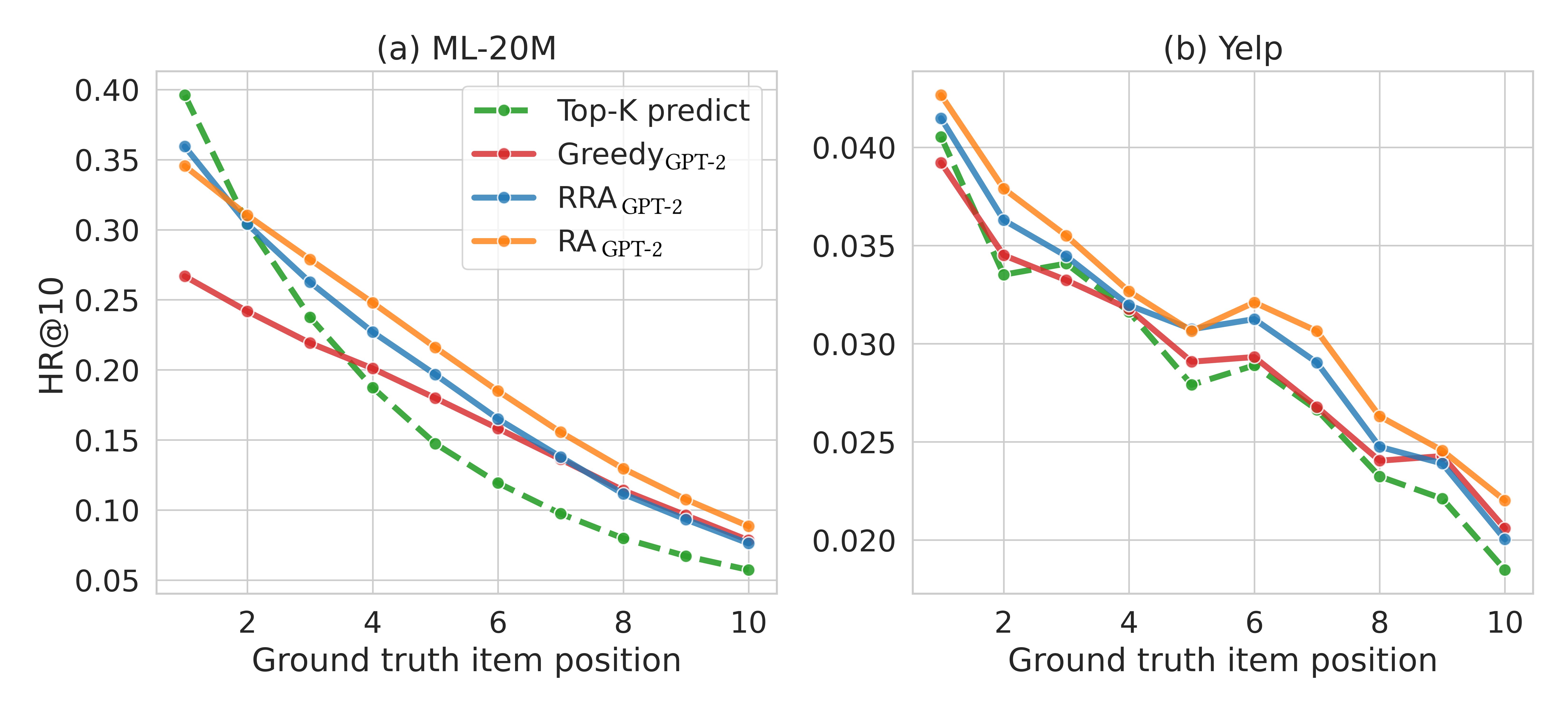}  
    \caption{Performance by ground truth item position on Movielens-20M and Yelp datasets. The dashed green line corresponds to the standard Top-K prediction strategy without generation.}
    \label{fig:position}
    \Description[Performance by ground truth item position]{Performance by ground truth item position}
\end{figure}

To more thoroughly evaluate long-term predictions, we have inspected the performance for each item position in test sequences separately. We consider only one item at a time from each user's sequence in the test set as ground truth and compare it with our recommendation list. Figure \ref{fig:position} shows how HitRate@10 (HR@10) depends on ground truth item position for MovieLens-20M and Yelp datasets. When looking at longer time horizons, performance degrades very quickly. However, degradation is less severe for recommendations made with the autoregressive generation strategies. Curves for MovieLens-20M are much smoother as this dataset is much bigger. The Top-K prediction strategy is better for the first positions, which is reasonable, as the model was directly trained to predict the next item in a sequence. For later positions, generation strategies outperform the Top-K prediction strategy. Our experimental evaluation confirms that autoregressive generation could be a valuable option for longer-term recommendations. Interestingly, greedy decoding and multi-sequence aggregation perform similarly for later positions on MovieLens-20M. The main difference is that aggregation is better for short-term predictions.

\section{Conclusion}
\label{sec:conclusion}
In this work, we explored different strategies for the autoregressive generation for the Top-K Sequential recommendation task. We found that commonly used single-sequence autoregressive generation strategies such as greedy, beam search, and temperature sampling do not outperform the Top-K prediction approach. However, the autoregressive generation strategies show higher performance in predicting longer-term user preferences.  Next, we found that greedy decoding outperforms temperature sampling and beam search. This observation contradicts experience in text generation, so we analyzed the reasons to explain this behavior.

Proposed multi-sequence aggregation approaches, \textit{Reciprocal Rank Aggregation strategy} and \textit{Relevance Aggregation strategy}, outperform single-sequence generation and Top-K prediction approaches. They require additional computational resources and increase inference time compared to single sequence generation approaches, but the inference time increase could be negligible due to parallelization.

The multi-sequence aggregation strategies may be used with various backbone models. Evaluation of proposed strategies with the other backbone models besides GPT-2 is one of the possible future research directions that can build upon this work. 


\bibliographystyle{ACM-Reference-Format}
\bibliography{recsys_content/7_bib}

\end{document}